\documentclass[aps,prd,preprint,groupedaddress,showpacs]{revtex4}
\usepackage{epsfig}
\usepackage{graphics}
\usepackage{slashed}
\usepackage{amssymb,amsmath}
\usepackage{color}
\begin{document}

\leftmargin -2cm
\def\choosen{\atopwithdelims..}

~~\\
 DESY~12--220 \hfill ISSN 0418-9833
\\November 2012

\vspace{2cm}

 \boldmath
\title{Drell-Yan lepton pair production at high energies in the Parton Reggeization Approach } \unboldmath

\author{\firstname{M.A. }\surname{Nefedov}}
\email{nefedovma@gmail.com}

\affiliation{Samara State University, Academician Pavlov Street 1,
443011 Samara, Russia}

\author{\firstname{N.N. }\surname{Nikolaev}}\email{n.nikolaev@fz-juelich.de}
\affiliation{ Institut f. Kernphysik, Forschungszentrum Juelich,
52425 Juelich, Germany}
 \affiliation{ L.D.Landau Institute for Theoretical Physics, Chernogolovka, 142432 Moscow Region,
Russia}

\author{\firstname{V.A.} \surname{Saleev}} \email{saleev@samsu.ru}
\affiliation{{II.} Institut f\"ur Theoretische Physik, Universit\"
at Hamburg, Luruper Chaussee 149, 22761 Hamburg, Germany}
\affiliation{ Samara State University, Academician Pavlov Street 1,
443011 Samara, Russia}

\begin{abstract}
According to extensive theoretical studies of the high energy limit
of QCD, inelastic interactions are dominated by the multi-Regge
final states. The appropriate gauge-invariant objects, which
simultaneously incorporate  the transverse momentum degrees of
freedom, are Reggeized gluons, quarks and antiquarks. In the present
communication we extend parton Reggeization approach to Drell-Yan
production of massive lepton pairs. The basic ingredient is a
process of Reggeized quark-antiquark annihilation, ${\cal Q}\bar
{\cal Q}\to \gamma^\star\to l^+l^-$, which  is described by the
Reggeon-Reggeon-photon effective vertex $\Gamma^{\gamma}_{{\cal
Q}\bar {\cal Q}}$. We calculate transverse-momentum and
invariant-mass distributions of Drell-Yan lepton pairs measured at
the CERN SPS, FNAL Tevatron and CERN LHC in the different ranges of
energy and rapidity. We focus on angular distributions of Drell-Yan
leptons in different kinematical ranges. The obtained results are
compared with the existing data and a good agreement is found. The
predictions for future experiments for Drell-Yan lepton pair
production at the CERN LHC  have been made.
\end{abstract}

\maketitle

\section{Introduction}
\label{sec:one}

The Drell-Yan production of lepton pairs, the principal QCD
subprocess for which is an annihilation of quarks and antiquarks
from colliding hadrons into lepton pairs \cite{DrellYan}, is not
only a supplement to the standard deep inelastic (DIS) studies.
Drell-Yan production stands out as an important source of the
experimental data on the flavor content of the nucleon sea (see
\cite{NuSea} and references therein). More recently there has been
much discussion of the Drell-Yan production as a unique source of
the direct information on fine spin properties of the proton,
including transversity and a new family of transverse-momentum
dependent (TMD) structure functions \cite{TMD1,TMD2}.

On the discovery frontier, Drell-Yan process emerges as a background
to heavy vector-meson production in the Standard Model (SM) and
beyond, as Drell-Yan lepton pairs with mass above the $Z-$bosom mass
are a major background in searches for new heavy vector mesons.
Finally, Drell-Yan production at the LHC would  probe  parton
distribution functions (PDFs) at very small $x$, down to $x\simeq
10^{-6}$. The experimental studies of Drell-Yan lepton pair
production include measurements of rapidity ($y$), invariant mass
($Q$), transverse momentum ($q_T=|{\vec q}_T|$) of lepton pairs, and
angular distributions of leptons in the virtual photon rest frame.

Theoretical study of Drell-Yan lepton pair production in the
perturbative Quantum Chromodynamics (pQCD) framework is in a very
advanced state and extends  from the leading-order (LO) to
next-to-leading-order (NLO) approximations in the strong coupling
constant $\alpha_s$ \cite{Stirling,Review,Berger1} to the soft
initial-state gluon resummation procedure to all orders in
$\alpha_s$   (see \cite{Resum,Berger2} and references therein).

Ever since the pioneering works by Dokshitser et al. \cite{DDT} and
Altarelli et al. \cite{Altarelli} there has been much work of the
$q_T-$distribution of lepton pairs in the framework of standard
collinear factorization with on-mass shell partons (see
\cite{Stirling,Review,Berger1,Resum,Berger2} and references
therein). More recent activity on $q_T-$distributions, which goes
beyond the collinear approximation, focused on TMD's \cite{Boer}.
The often discussed $k_T-$factorization with off-shell mass partons
is a part of such approach. At small-x of our interest the principal
ingredients are off-shell properties of $t-$channel exchanges and
unpolarized unintegrated parton distribution functions
(PDF)\cite{KT1,KT2}. Indeed, at a deeper level of high energy pQCD,
quarks, antiquarks and gluons are known to reggeize \cite{Lipatov95,
BFKL, FadinSherman, FadinLipatov}. Furthermore, as has been shown by
Lipatov and collaborators, Reggeized gluons and quarks are the
appropriate gauge-invariant degrees of freedom of high-energy pQCD.
In the practical applications, the use of Reggeized $t-$channel
exchanges is justified by the dominance of the so-called multi-Regge
final states in inelastic collisions of high energy hadrons.

Earlier, the quark Reggeization hypothesis has been used
successfully for a description of different spectra of prompt
photons at the Fermilab Tevatron and CERN LHC
\cite{tevatronY,tevatronYY,lhcY}, electron deep inelastic scattering
and prompt photon production cross sections at the DESY HERA
\cite{tevatronY,heraY}, and forward $Z-$boson production cross
section at the CERN LHC \cite{HautmanZ}. In the present
communication we discuss the experimental data for proton-proton and
proton-antiproton collisions including the Fermilab Tevatron
experiments \cite{DY_CDF,DY_D0, NuSea} and the CERN SPS experiments
\cite{R209,UA1}, and new data from the CERN LHC \cite{CMSDY,LHCbDY}.

This paper is organized as follows: In Sec.~II, the general
formalism for description of Drell-Yan lepton pair production is
presented. In Sec.~III, we calculate helicity structure functions in
the LO of Parton Reggeization Approach (PRA) and obtain master
formulas for lepton pair spectra and angular coefficients. In
Sec.~IV, we present results of our calculations for invariant-mass
and transverse-momentum spectra of Drell-Yan lepton pairs and report
a comparison with relevant experimental data from the Fermilab
Tevatron, CERN SPS and CERN LHC. We also compare our predictions
with the data for angular coefficients and predict
transverse-momentum dependence of the angular coefficients in the
energy range of the CERN LHC, $\sqrt{S}=7-14$ TeV. Section V
contains our conclusions.

\section{Drell-Yan pair production in QCD and PRA}
\label{sec:two}

According to the perturbative  QCD  Parton Model the LO subprocess
of Drell-Yan lepton pair production is an annihilation of quark and
antiquark from the colliding hadrons into the virtual photon which
decays onto lepton pair:
\begin{equation}
q+\bar q \to \gamma^\star \to l^+ + l^-.\label{qqDY}
\end{equation}
In this formalism the quark and antiquark in an initial state have
zero transverse momentum with respect to the hadron collision axis
and massive lepton pair (or virtual photon) is produced with zero
transverse momentum. The angular distribution of leptons in the
virtual photon rest frame should be
\begin{equation}\frac{dN}{d\cos\theta}\sim 1+\cos^2\theta\label{cos2},\end{equation}
where $\theta$ is the polar angle of the lepton relative to the
collision axis.  Experimentally, massive lepton pairs are produced
with substantial transverse momentum ${q}_T$. To describe Drell-Yan
lepton pairs  with nonzero transverse momentum, within the collinear
Parton Model one invokes the NLO partonic subprocesses $2\to 2$,
which are of the first order in $\alpha_s$:
\begin{eqnarray}
q+\bar q\to g + \gamma^\star\to g+l^++ l^-,\label{ann}\\
q+g\to q+\gamma^\star\to q+l^+ +l^-\label{comp},
\end{eqnarray}
where $g$ is the Yang-Mills gluon. It is obvious, that in these
subprocesses the lepton angular distribution will differ from the
trivial form  (\ref{cos2}). If the colliding hadrons were polarized,
their polarization transfer to the initial partons and angular
distributions of the final leptons will change respectively. In
framework of the NLO QCD and the collinear Parton Model, the
Drell-Yan lepton pair production in collisions of polarized and
unpolarized hadrons have been studied carefully, with exception of
only the regions of small $q_T$ and $Q$
\cite{Stirling,Review,Berger1}.

In the $k_T-$factorization approach \cite{KT1,KT2},  the off-shell
initial partons $(q^*, g^*)$ of nonzero transverse momenta
 are considered from the beginning. Instead of processes (\ref{ann})
and (\ref{comp}), the relevant LO and NLO contributions are
\begin{eqnarray}
q^*+\bar q^*\to  \gamma^\star\to l^++ l^-,\label{annKT0}\\
q^*+\bar q^*\to g + \gamma^\star\to g+l^++ l^-,\label{annKT}\\
q^*+g^*\to q+\gamma^\star\to q+l^+ +l^-\label{compKT}.
\end{eqnarray}
However, one must be careful with off-shell quarks and antiquarks as
one may break the gauge invariance of relevant amplitudes and break
the electromagnetic  current conservation.  The additional problem
of the $k_T-$factorization calculations is a double counting, when
 subprocesses (\ref{annKT0}), (\ref{annKT}) and (\ref{compKT}) are taken into
account together. These difficulties can be solved  in PRA, where
the initial off-shell gluons and quarks are considered as Reggeons
or Reggeized gluons and quarks, which interact with usual quarks and
Yang-Mills gluons in a special way, via gauge invariant effective
vertices which incorporate the initial and final state radiation
effects on equal {footing}
\cite{FadinLipatov,FadinSherman,Antonov,LipatoVyazovsky}. Our
previous studies of inclusive jet \cite{lhcY}, and inclusive prompt
photon production \cite{tevatronY,heraY} have shown that in PRA it
is sufficient to consider only LO $2\to 1$ subprocess for a good
quantitative description the experimental data.

In PRA, the LO subprocess, which describes finite $q_T$ Drell-Yan
lepton pairs, is an annihilation of Reggeized quark and Reggeized
antiquark via virtual photon:
\begin{equation}
{\cal Q}(q_1) +\bar{\cal Q}(q_2)\to \gamma^\star\to l^+(k_1)  +
l^-(k_2)\label{QQDY}.
\end{equation}
The amplitude of the subprocess (\ref{QQDY}) reads as follows
\begin{equation}
M({\cal Q}_i\bar{\cal Q}_i\to l^+l^-) = 4\pi \alpha e_i \bar
V(x_2P_2)\Gamma^{\gamma,\mu}_{{\cal Q}\bar{\cal
Q}}(q_1,q_2)U(x_1P_1) \otimes \bar U(k_1)\gamma_\mu
V(k_2),\label{QQgamma}
\end{equation}
where $e_i$ is the electric charge of quark $i$ (in units of
electron charge), $\alpha$ is the electromagnetic constant, and
$\Gamma^{\gamma,\mu}_{{\cal Q}\bar{\cal Q}}(q_1,q_2)$ is the
Fadin-Sherman effective vertex \cite{FadinSherman,LipatoVyazovsky},
\begin{equation}
\Gamma^{\gamma,\mu}_{{\cal Q}\bar{\cal
Q}}(q_1,q_2)=\gamma^\mu-\frac{2\hat q_1P_1^\mu}{x_2S}-\frac{2\hat
q_2P_2^\mu}{x_1S}.
\end{equation}
It is easy to show that the amplitude (\ref{QQgamma}) is gauge
invariant and  $\Gamma^{\gamma,\mu}_{{\cal Q}\bar{\cal
Q}}(q_1,q_2)(q_1+q_2)_\mu\equiv 0$. Four-momenta of Reggeized quarks
(antiquarks) have transverse components and they read
$q_i^\mu=x_iP_i^\mu+q_{iT}^\mu$, $q_{iT}^\mu=(0,\vec q_{iT},0)$,
$q_i^2=q_{iT}^2=-\vec q_{iT}^{~2}=-t_i\neq 0$.

The Drell-Yan pair production in the proton-proton and
proton-antiproton high-energy collisions corresponds to the
following processes
\begin{eqnarray}
p(P_1)+ \bar p(P_2) \to l^+(k_1)+ l^-(k_2) + X,\label{TEV}\\
p(P_1)+ p(P_2) \to l^+(k_1)+ l^-(k_2) + X,\label{LHC}
\end{eqnarray}
where four-momenta of particles are shown in brackets, $l=e,\mu$
(electron or muon), $q=q_1+q_2=k_1+k_2$  is the four-momentum of
virtual photon,  $Q=\sqrt{q^2}$ and $Q_T^2=Q^2+{\vec
q}_T^{~2}=x_1x_2S$.  Differential cross section for  processes
(\ref{TEV}) and (\ref{LHC}) have the standard form:
\begin{eqnarray}
\frac{d\sigma}{d^4q
d\Omega}=\frac{\alpha^2}{32\pi^4SQ^4}L_{\mu\nu}W^{\mu\nu}\label{sech00}
\end{eqnarray}
or
\begin{eqnarray}
\frac{d\sigma}{dQ^2dq_T^2dyd\Omega}=\frac{\alpha^2}{64\pi^3SQ^4}L_{\mu\nu}W^{\mu\nu},\label{sech0}
\end{eqnarray}
where $y$ is the rapidity of virtual photon (or $l^+l^-$ lepton
pair), $d\Omega=d\phi d\cos\theta$ is the spatial  angle of
producing positive lepton in the rest frame of virtual photon,
$P_1=\frac{\sqrt{S}}{2}(1,0,0,1)$,
$P_2=\frac{\sqrt{S}}{2}(1,0,0,-1)$, $\sqrt{S}$ is the total energy
of colliding particles. Here
\begin{equation} L^{\mu\nu}=2(k_1^\mu
k_2^\nu+k_1^\nu k_2^\mu)-Q^2 g^{\mu\nu}\end{equation} is the
leptonic tensor, whereas
\begin{equation} W_{\mu\nu}=\int d^4 x
e^{iqx}\langle P_1P_2|j_\mu(x)j_\nu(0)|P_1P_2 \rangle\end{equation}
is the hadronic tensor.

The convolution of hadronic and leptonic tensors reads as a sum of
contributions of the so-called helicity structure functions
$W_{T,L,\Delta,\Delta\Delta}$ \cite{Berger2,helicityDY}:
\begin{eqnarray}
\frac{d\sigma}{dQ^2dq_T^2dyd\Omega}&=&\frac{\alpha^2}{64\pi^3SQ^2}\Bigl[W_{T}(1+\cos^2\theta)
+W_L(1-\cos^2\theta)+\nonumber\\
&+&W_{\Delta}\sin 2\theta\cos\phi+ W_{\Delta\Delta}\sin^2\theta\cos
2\phi\Bigr].\label{sech1}
\end{eqnarray}
After integration over the angles $\theta$ and $\phi$ in
(\ref{sech1}), we obtain
\begin{eqnarray}
\frac{d\sigma}{dQ^2dq_T^2dy}&=&\frac{\alpha^2}{64\pi^3SQ^2}\left(\frac{16\pi}{3}\right)
W_{TL},\label{sech2}
\end{eqnarray}
where $W_{TL}=W_T+W_L/2$. In the analysis of experimental data, the
angular distribution of leptons is represented in terms of two sets
of the angular coefficients:
\begin{eqnarray}
\frac{dN}{d\Omega}&=&(1+\cos^2\theta)+A_0\left(
\frac{1}{2}-\frac{3}{2}\cos^2\theta \right)+A_1\sin
2\theta\cos\phi+\frac{A_2}{2}\sin^2\theta\cos 2\phi
\end{eqnarray}
and
\begin{eqnarray}
\frac{dN}{d\Omega}&=& \frac{4}{\lambda+3}\left(1+\lambda
\cos^2\theta+\mu\sin 2\theta\cos\phi+\frac{\nu}{2}\sin^2\theta\cos
2\phi\right),
\end{eqnarray}
with the normalization
\begin{eqnarray}
\int \left(\frac{dN}{d\Omega}\right)d\Omega=\frac{16 \pi}{3}.
\end{eqnarray}

 One set consists of the coefficients
\begin{eqnarray}
A_0=\frac{W_L}{W_{TL}}, \quad A_1=\frac{W_\Delta}{W_{TL}}, \quad
A_2=\frac{2W_{\Delta\Delta}}{W_{TL}}\label{AAA},
\end{eqnarray}
and the other one is defined by
\begin{eqnarray}
\lambda=\frac{2-3A_0}{2+A_0}, \quad \mu=\frac{2A_1}{2+A_0}, \quad
\nu=\frac{2A_2}{2+A_0}.
\end{eqnarray}
Helicity structure functions are obtained by the projection of
hadronic tensor on the photon states with the different
polarizations $\epsilon_\lambda^\mu(q)$,  $\lambda=\pm 1,0$:
\begin{eqnarray}
W_T&=&W_{\mu\nu}\epsilon_{+1}^{\mu\star}\epsilon_{+1}^\nu,\\
W_L&=&W_{\mu\nu}\epsilon_{0}^{\mu\star}\epsilon_{0}^\nu,\\
W_\Delta&=&W_{\mu\nu}\left(\epsilon_{+1}^{\mu\star}\epsilon_{0}^\nu+
\epsilon_{0}^{\mu\star}\epsilon_{+1}^\nu\right)/\sqrt{2},\\
W_{\Delta\Delta}&=&W_{\mu\nu}\epsilon_{+1}^{\mu\star}\epsilon_{-1}^\nu.
\end{eqnarray}
In the reference frame of the virtual photon its polarization
4-vector can be written in covariant form:
\begin{equation}
\epsilon_{\pm 1}^\mu=\frac{1}{\sqrt{2}}(\mp X^\mu-i Y^\mu), \quad
\epsilon_0^\mu=Z^\mu,
\end{equation}
where 4-vectors $X,Y,Z$ satisfy following conditions:
$X^2=Y^2=Z^2=-1$, $q_\mu X^\mu=q_\mu Y^\mu=q_\mu Z^\mu=0$. In the
Collins-Soper frame \cite{CollinsSoper}, in which we are working,
these unit vectors are defined as:
\begin{eqnarray}
Z^\mu&=&\frac{2}{Q_T\sqrt{S}}\left[ (qP_2)\tilde
P_1^\mu-(qP_1)\tilde P_2^\mu\right],\\
X^\mu&=&-\frac{2Q}{q_TQ_T\sqrt{S}}\left[ (qP_2)\tilde
P_1^\mu+(qP_1)\tilde P_2^\mu\right],\\
Y^\mu&=&\varepsilon^{\mu\nu\alpha\beta}T_\nu Z_\alpha X_\beta,
\end{eqnarray}
where $$T^\nu=\frac{q^\nu}{Q}, \quad \tilde
P_i^\mu=\frac{1}{\sqrt{S}}\bigl(
P_i^\mu-\frac{(qP_i)}{Q^2}q^\mu\bigr).$$

\section{Helicity structure functions in PRA}

To calculate components of the hadronic tensor $W^{\mu\nu}$  and
helicity structure functions $W_{T,L,\Delta,\Delta\Delta}$ we need
to know squared modula of partonic amplitudes and relevant PDFs from
colliding hadrons. In the processes of massive lepton pair
production with large transverse momenta, where $Q, q_T
>>\Lambda_{QCD}\simeq 0.1-0.2$ GeV, one has a factorization of hard scattering subprocesses
at the scale $\mu \sim Q_T$ and the perturbative QCD evolution of
the parton distributions from some initial scale $\mu_0\simeq 1$ GeV
to the hard scattering scale $\mu$. In the collinear Parton Model
hadronic and partonic cross sections are connected by the textbook
factorization formula:
\begin{equation}
d\sigma (pp\to l^+l^-X)=\sum_{q}\int dx_1\int dx_2
f^p_q(x_1,\mu^2)f^p_{\bar q}(x_2,\mu^2) d\hat\sigma(q\bar q\to
l^+l^-)\label{DYpm},
\end{equation}
where $f^p_q(x_{1,2},\mu^2)$ is quark (antiquark) collinear PDFs.
The PRA generalization of the same factorization formula,
differential in the Reggeized partons virtualities, is
\begin{eqnarray}
d\sigma (pp\to l^+l^-X)&=&\sum_{q}\int \frac{d\phi_1}{2\pi}\int
dt_1\int \frac{dx_1}{x_1}\int \frac{d\phi_2}{2\pi}\int dt_2\int
\frac{dx_2}{x_2}\nonumber\\
 &&
 \Phi^p_q(x_1,t_1,\mu^2)\Phi^p_{\bar q}(x_2,t_2,\mu^2)
d\hat\sigma({\cal Q}\bar {\cal Q}\to l^+l^-)\label{DYkt}.
\end{eqnarray}
The unintegrated PDFs $\Phi_q^p(x,t,\mu^2)$ are related to their
collinear counterparts $f_q^p(x,\mu^2)$ by the normalization
condition
\begin{equation}
xf_q^p(x,\mu^2)=\int^{\mu^2}dt\,\Phi_q^p(x,t,\mu^2),
\end{equation}
which furnishes a correct transition from formulas in PRA to those
in the collinear Parton Model.  In our numerical analysis, we adopt
the prescription proposed by Kimber, Martin, and Ryskin (KMR)
\cite{KMR} to obtain the unintegrated quark PDFs of the proton from
the conventional integrated one. Factorization scale is chosen to be
$\mu=\xi Q_T$, and varying $\xi$  between $1/2$ and 2 serves as an
estimate of the scale uncertainty. It was found, that magnitudes of
angular observables   ($A_0,\ A_2,\ \lambda,\ \nu$) are much more
stable under the scale variations than  differential cross sections.
In the kinematical regions under consideration, their variations
under scale change are found to be below  20 \%, and indicated as
shaded bands in the figures with theoretical predictions.

Partonic cross sections are connected with squared amplitudes of
subprocesses in both the collinear Parton Model and  PRA  the
standard way:
\begin{eqnarray}
d\sigma (q({\cal Q})\bar q(\bar {\cal Q})\to
l^+l^-)&=&(2\pi)^4\delta^{(4)}(q_1+q_2-k_1-k_2)\frac{\overline{|M(q({\cal
Q})\bar q(\bar {\cal Q})\to l^+l^-)|^2}}{2x_1x_2S}\times \nonumber\\
&&\times \frac{d^3k_1}{(2\pi)^3 2k_{10}}\frac{d^3k_2}{(2\pi)^3
2k_{20}}.
\end{eqnarray}
Taking into account that
\begin{equation}
\int \frac{d^3k_1}{k_{10}}\int \frac{d^3k_2}{k_{20}}=\frac{1}{2}\int
d^4q\int d\Omega,
\end{equation}
we obtain
\begin{equation}
d\sigma (q({\cal Q})\bar q(\bar {\cal Q})\to
l^+l^-)=\delta^{(4)}(q_1+q_2-q)\frac{\overline{|M(q({\cal Q})\bar
q(\bar {\cal Q})\to l^+l^-)|^2}}{x_1x_2S}\frac{d^4q
d\Omega}{64\pi^2}.
\end{equation}

In the collinear Parton Model the well-known answer is
\begin{equation}
\overline{|M(q_i\bar q_i\to
l^+l^-)|^2}=\frac{16\pi^2}{3Q^4}\alpha^2e_i^2
L^{\mu\nu}w_{\mu\nu}^{PM},
\end{equation}
where \begin{equation}w_{\mu\nu}^{PM}=x_1x_2(2P_1^\mu
P_2^\nu+2P_1^\nu P_2^\mu-S g^{\mu\nu})\label{ptensor}\end{equation}
is the quark tensor, in which it is supposed that $q_i^\mu=x_i
P_i^\mu$.

On the other hand, in  PRA we obtain for the squared amplitude of
the subprocess (\ref{QQDY}):
\begin{equation}
\overline{|M({\cal Q}_i\bar {\cal Q}_i\to
l^+l^-)|^2}=\frac{16\pi^2}{3Q^4}\alpha^2e_i^2
L^{\mu\nu}w_{\mu\nu}^{PRA},
\end{equation}
where the tensor of Reggeized quarks reads:
\begin{eqnarray}
w_{\mu\nu}^{PRA}&=&x_1x_2\bigl[-Sg^{\mu\nu}+2(P_1^\mu P_2^\nu+
P_2^\mu
P_1^\nu)\frac{(2x_1x_2S-Q^2-t_1-t_2)}{x_1x_2S}+\nonumber\\
&+&\frac{2}{x_2}(q_1^\mu P_1^\nu+q_1^\nu
P_1^\mu)+\frac{2}{x_1}(q_2^\mu P_2^\nu+q_2^\nu P_2^\mu)+ \label{Qtensor} \\
 &+&\frac{4(t_1-x_1x_2S)}{Sx_2^2}P_1^\mu P_1^\nu+
\frac{4(t_2-x_1x_2S)}{Sx_1^2}P_2^\mu P_2^\nu\bigr].\nonumber
\end{eqnarray}
Note, the both tensors (\ref{ptensor}) and (\ref{Qtensor}) satisfy
gauge-invariance condition $q^\mu w_{\mu\nu}^{PRA}=q^\mu
w_{\mu\nu}^{PM}=0$. It is obvious that $w_{\mu\nu}^{PRA}\to
w_{\mu\nu}^{PM}$, if we put $t_i\to 0$ and $q_i^\mu\to x_iP_i^\mu$.

It is interesting to define the quark helicity structure functions
$w_{T,L,\Delta,\Delta\Delta}$ respectively to the hadron helicity
structure functions  $W_{T,L,\Delta,\Delta\Delta}$. Upon direct
calculations we obtain
\begin{equation}
w_T^{PM}=Q^2,\quad w_L^{PM}=\quad w_\Delta^{PM}=\quad
w_{\Delta\Delta}^{PM}=0,
\end{equation}
whereas
\begin{eqnarray}
w_T^{PRA}&=&Q^2+\frac{(\vec q_{1T}+\vec q_{2T})^2}{2}, \quad
w_L^{PRA}=(\vec q_{1T}-\vec q_{2T})^2 \label{wTLPRA}\\
w_\Delta^{PRA}&=&0,\quad w_{\Delta\Delta}^{PRA}=\frac{(\vec
q_{1T}+\vec q_{2T})^2}{2}.\label{wDPRA}
\end{eqnarray}
We  can also calculate the partonic angular coefficients, which are
defined like the hadronic angular coefficients (\ref{AAA}):
\begin{eqnarray}
a_0=\frac{{w}_{L}^{PRA}}{w_{TL}^{PRA}}=\frac{(\vec q_{1T}-\vec
q_{2T})^2}{Q^2+t_1+t_2},\quad
a_1=\frac{w_{\Delta}^{PRA}}{w_{TL}^{PRA}}=0,\quad a_2=\frac{2
w_{\Delta\Delta}^{PRA}}{w_{TL}^{PRA}}=\frac{(\vec q_{1T}+\vec
q_{2T})^2}{Q^2+t_1+t_2}.
\end{eqnarray}
Upon averaging over the angle, $\phi=\phi_1-\phi_2$, between
transverse momenta of the Reggeized quarks, we obtain well-known
Lam-Tung \cite{LamTung} relation for the partonic angular
coefficients
\begin{equation}
<a_0>_{\phi}= <a_2>_{\phi}.\label{eq:LamTung}
\end{equation}
This relation, as it will be shown below, breaks  strongly for the
hadronic angular coefficients $A_0$ and $A_2$, when $Q$ and $q_T$
become smaller at the fixed collision energy.

The helicity structure functions $W_{T,...}^{PRA}$ at the fixed
values of variables  $S,Q,q_T,y$ can be presented via corresponding
quark helicity functions $w^{PRA}_{T,...}$:
\begin{equation}
W_{T,...}^{PRA}(S,Q,q_T,y)=\frac{8\pi^2S}{3Q_T^4}\int dt_1\int
d\phi_1 \sum_q \Phi_q^p(x_1,t_1,\mu^2)\Phi_{\bar q}^p(x_2,t_2,\mu^2)
w^{PRA}_{T,...}. \label{WPRA}
\end{equation}

The experimental data for sets of angular coefficients $A_0,A_1,A_2$
or $\lambda,\mu,\nu$ are presented as averages  over certain
kinematical region of variables  $Q,q_T,y$ at the fixed value of
collision energy $\sqrt{S}$. We define the relevant averages  as
follows
\begin{eqnarray}
\overline{W^{regge}_{T,...}(S)}=J^{-1}\int_{Q_{min}}^{Q_{max}}
dQ\int_{q_{Tmin}}^{q_{Tmax}} dq_T\int_{y_{min}}^{y_{max}} dy
W_{T,...}^{regge}(S,Q,q_T,y),
\end{eqnarray}
where $J=(Q_{max}-Q_{min})(q_{Tmax}-q_{Tmin})(y_{max}-y_{min})$.
Sometimes instead of a variable $y$ one uses the Feynman variable
$x_F$, which is defined in the center of mass frame of colliding
hadrons as follows
\begin{equation}
x_F=\frac{2
q_z}{\sqrt{S}}=\frac{Q_T}{\sqrt{S}}\left(e^y-e^{-y}\right),\label{fxf}
\end{equation}
and
\begin{equation}
dy=\frac{dx_F}{x_F\sqrt{1+\frac{4Q_T^2}{Sx_F^2}}}.
\end{equation}
In other cases one uses the pseudorapidity
\begin{equation}
\eta=\frac{1}{2}\ln \frac{|\vec q|+q_z}{|\vec q|-q_z}=\frac{1}{2}\ln
\left(\frac{1+\cos\theta}{1-\cos\theta}\right), \quad dy=\frac{|\vec
q|}{q_0} d\eta,\label{feta}
\end{equation}
where
\begin{eqnarray}
\cos\theta=\frac{e^{2\eta}-1}{e^{2\eta}+1},\quad |\vec
q|=\frac{q_T}{|\sin\theta|},\quad q_0=\sqrt{Q^2+|\vec q|^2}.
\end{eqnarray}

Our LO PRA results {for the cross-sections} should be corrected by
the so-called K-factor, which includes high order (HO) QCD
corrections to the LO diagrams. The main part of HO corrections
arising from real gluon emission is already accounted in LO PRA.
Still another part comes from the non-logarithmic loop corrections
arising from gluon vertex corrections. Accordingly to
Ref.~\cite{Kfactor}, this K-factor is written as follows
\begin{equation}
K({\cal Q}\bar{\cal Q}\to
\gamma^*)=\exp(C_F\frac{\alpha_s(\mu^2)}{2\pi}\pi^2),\label{Kfact}
\end{equation}
where a particular scale choice $\mu^2=Q_T^{4/3}Q^{2/3}$ for
evaluation $\alpha_s(\mu^2)$ in (\ref{Kfact}) has been advocated in
\cite{Kfactor}. This phenomenological Anzatz can be used for large
$Q^2$ Drell-Yan lepton pair production only. In case of quasi-real
photon production, when $q_T\gg Q\sim \Lambda_{QCD}$, we put
$K({\cal Q}\bar{\cal Q})=1$ in accordance with calculations for the
real photon production \cite{lhcY}. The typical numerical value of
K-factor at the kinematical conditions under consideration is about
$1.3\sim 1.8$.

\section{Results}

\begin{figure}[p]
\begin{center}
\includegraphics[width=.6\textwidth, clip=]{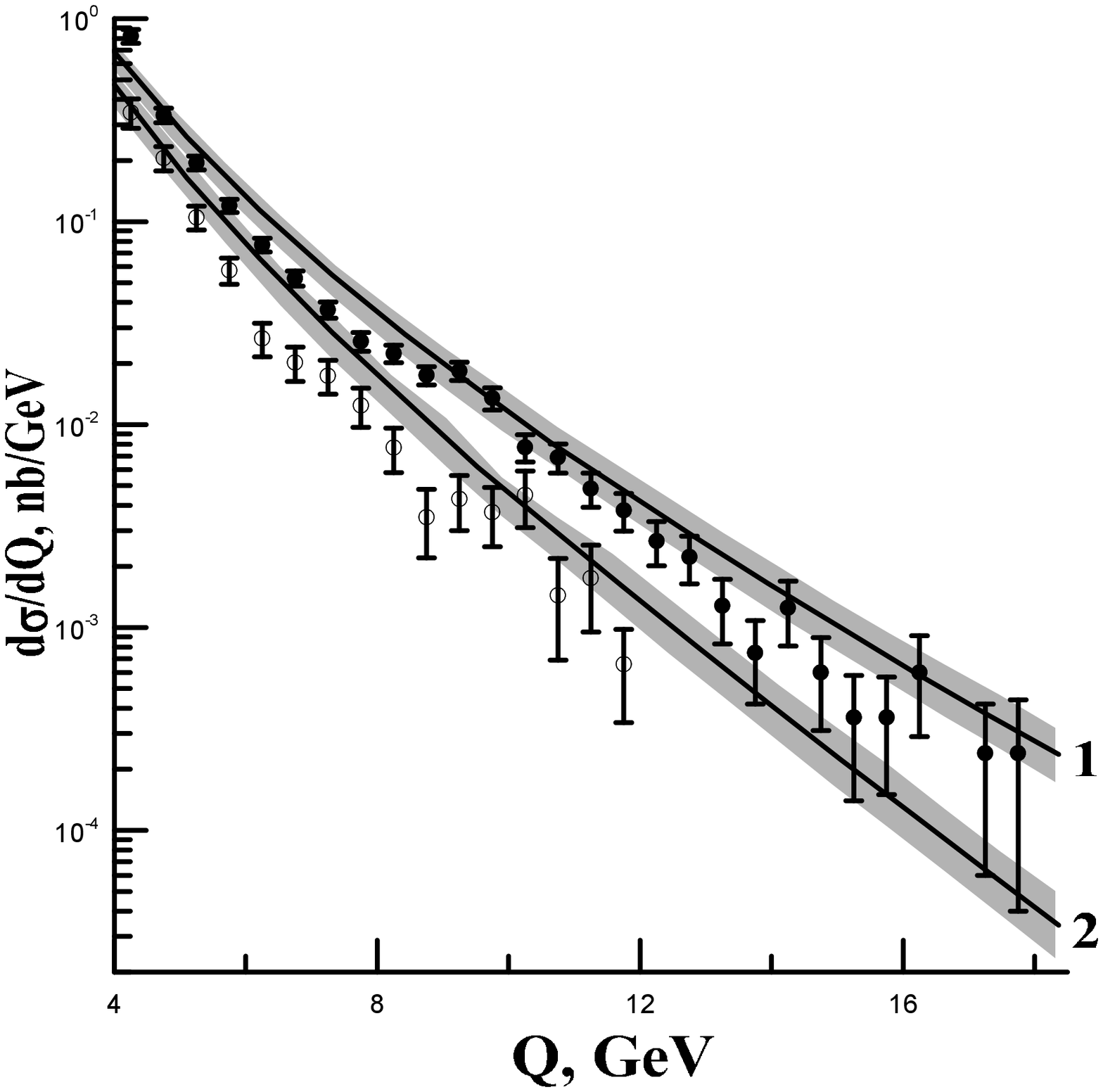}
\end{center}
\caption{Differential cross section of Drell-Yan lepton pair
production as function of virtual photon mass $Q$. The data are from
R209 Collaboration \cite{R209}. The curve 1 -- $\sqrt{S}=62$ GeV,
the curve 2 -- $\sqrt{S}=44$ GeV.}\label{R209Q}
\end{figure}

\begin{figure}[p]
\begin{center}
\includegraphics[width=.6\textwidth, clip=]{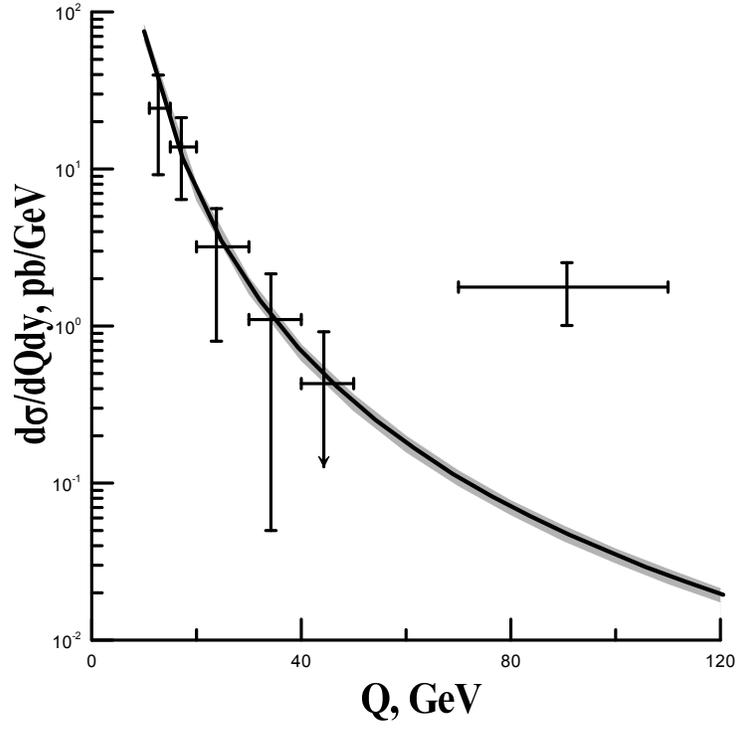}
\end{center}
\caption{Doubly differential cross section of Drell-Yan lepton pair
production as function of virtual photon mass $Q$. The data are from
CDF Collaboration \cite{DY_CDF} at $\sqrt{S}=1.8$ TeV.}\label{CDFQ}
\end{figure}

\begin{figure}[p]
\begin{center}
\includegraphics[width=.6\textwidth, clip=]{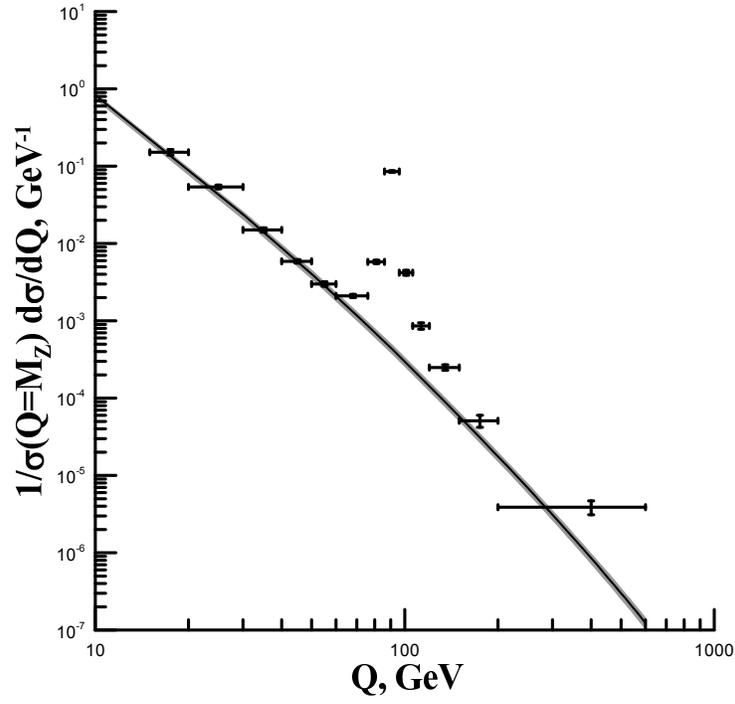}
\end{center}
\caption{Differential cross section of Drell-Yan lepton pair
production as function of virtual photon mass $Q$. The data are from
CMS Collaboration \cite{CMSDY} at $\sqrt{S}=7$ TeV. Solid line - our
prediction normalized to the theoretical value of the cross-cection
in the Z-boson region.}\label{CMSQ}
\end{figure}

 We start our comparison of theoretical predictions with the experimental data for the
invariant-mass distributions of the Drell-Yan lepton pairs. In
Fig.~\ref{R209Q}, the predictions of the LO PRA are compared with
the data from the R209 Collaboration at the two values of collision
energy $\sqrt{S}=44$ GeV and $\sqrt{S}=62$ GeV. The resulting
invariant-mass spectrum has been obtained by integration over the
 {relevant} virtual photon transverse momentum {range} and rapidity range $|y|<4$. In Fig.~\ref{CDFQ}, the doubly
differential cross section $d\sigma/dQdy$ is presented as function
of $Q$ at the $\sqrt{S}=1.8$ TeV, $|y|<1$ and $q_T<200$ GeV as it
has been measured by CDF Collaboration \cite{DY_CDF}. In
Fig.~\ref{CMSQ}, the data from CMS Collaboration \cite{CMSDY} are
presented as $Q-$spectrum normalized to the cross section in the
$Z-$boson peak. {The solid line represents our prediction,
normalized on the value of cross section in the Z-boson region
($\sigma(60<Q<120\mbox{ GeV})=973$ pb), obtained theoretically in
the Ref. \cite{CMSDY}, see  Table 10 therein.}

As one would expect, the increase in energy improves an agreement
between the theory and experiment. Of course, the $Z-$boson region,
$60\leq Q \leq 120$ GeV, is beyond our consideration. An extension
of PRA to the region of the $Z-$boson production demands a use of
the unknown up today effective vertex $\Gamma^{Z}_{{\cal Q}\bar
{\cal Q}}$ which describes the  Reggeized quark-antiquark
annihilation in $Z-$boson.

\begin{figure}[p]
\begin{center}
\includegraphics[width=.6\textwidth, clip=]{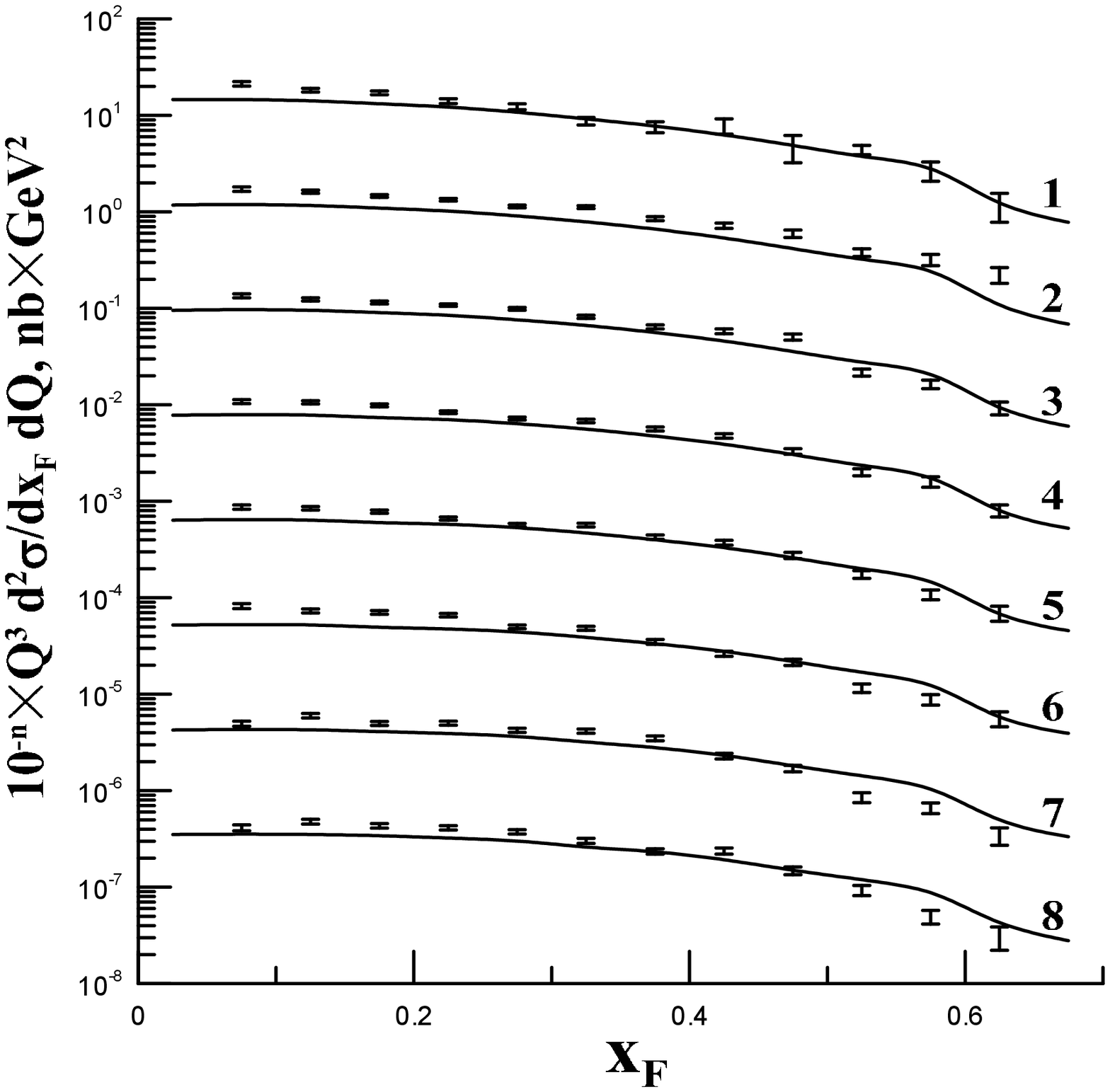}
\end{center}
\caption{Differential cross section $Q^3 d\sigma/dx_F dQ$ of
Drell-Yan lepton pair production as function of $x_F$ integrated
over all $q_T$. Curves $1-8$ correspond $Q$ from $4.75$ GeV till
$Q=8.25$ GeV with the step equal to $0.5$ GeV. The data are from
FNAL fixed target experiment \cite{FNALXF} at the $\sqrt{S}=38.8$
GeV.} \label{FNALXF}
\end{figure}

To demonstrate an agreement between PRA and Drell-Yan pair spectra
over the longitudinal variables, we show in the Fig.~\ref{FNALXF}
the differential cross section $Q^3 d\sigma/dx_F dQ$  as a function
of the Feynman variable $x_F$ after integration over all $q_T$.
Curves $1-8$ correspond to $Q$ from $4.75$ GeV till $Q=8.25$ GeV
with the step of $0.5$ GeV. The data are from FNAL fixed target
experiment \cite{FNALXF} at the $\sqrt{S}=38.8$ GeV.

\begin{figure}[ht]
\begin{center}
\includegraphics[width=.6\textwidth, clip=]{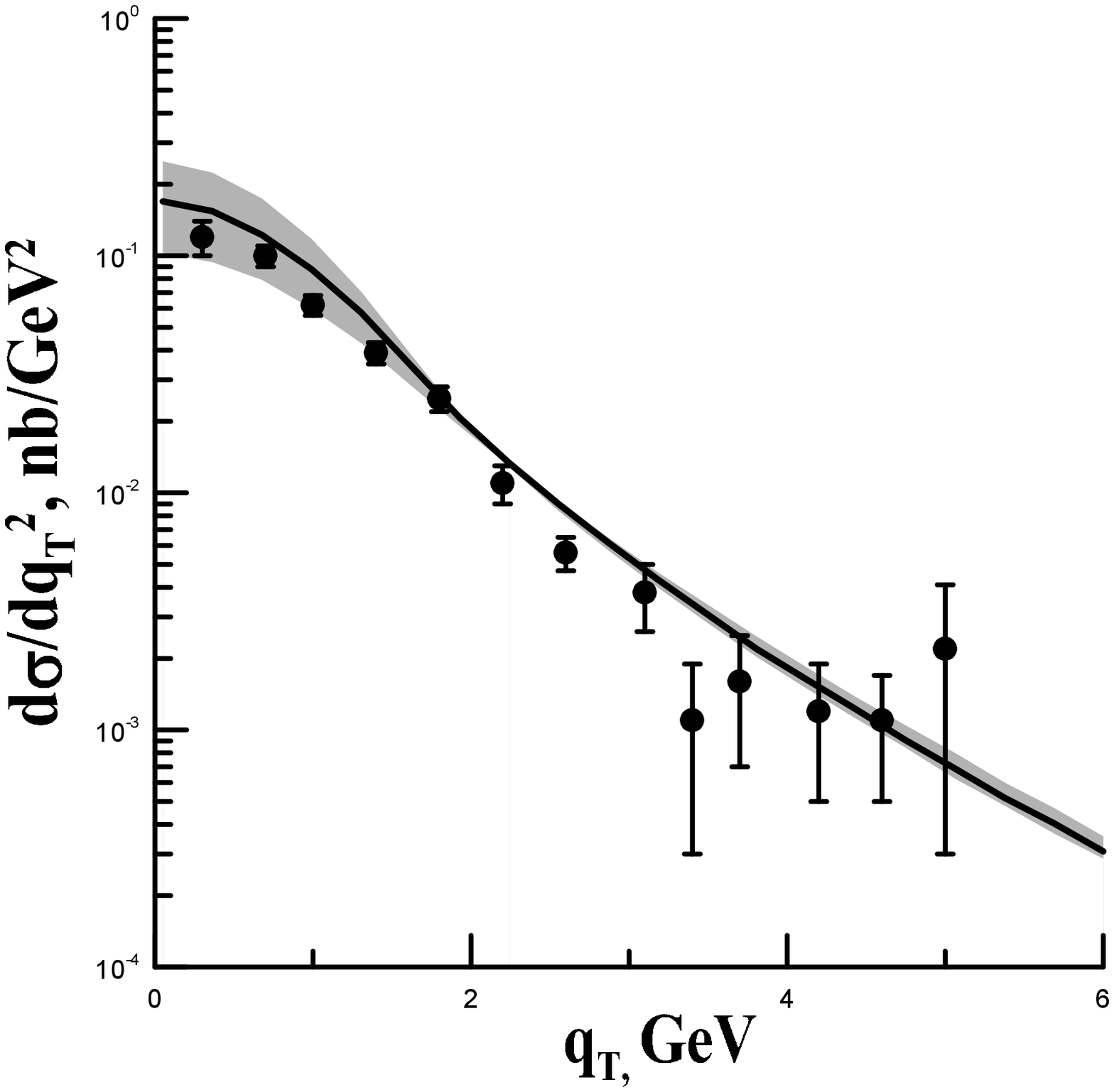}
\end{center}
\caption{ Differential cross section of Drell-Yan lepton pair
production as function of $q_T$.  The data are from  R209
Collaboration \cite{R209} at the $|y|<4$, $5<Q<8$ GeV, $\sqrt{S}=62$
GeV.}\label{R209qt}
\end{figure}

\begin{figure}[ht]
\begin{center}
\includegraphics[width=.6\textwidth, clip=]{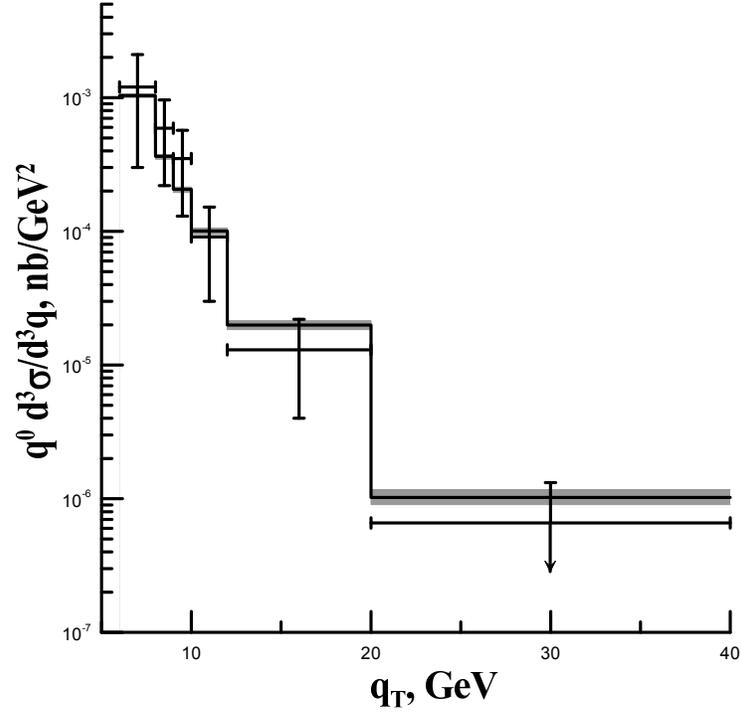}
\end{center}
\caption{ Differential cross section of Drell-Yan lepton pair
production as function of $q_T$.  The data are from  UA1
Collaboration \cite{UA1} at the $|y|<1.7$, $0.2<Q<2.5$ GeV,
$\sqrt{S}=630$ GeV.}\label{UA1qt}
\end{figure}

The transverse-momentum dependence of Drell-Yan lepton pair
production cross section is demonstrated in  Figs.~\ref{R209qt} and
\ref{UA1qt}. The R209 Collaboration \cite{R209} has measured
$q_T-$spectrum at $\sqrt{S}=62$ GeV, $|y|<4$, and $5<Q<8$ GeV. We
describe this data quite well, especially at  small transverse
momenta $q_T<2$ GeV, where the pure  NLO collinear Parton Model
calculations break down and where one usually invokes unknown
(ad'hoc) nonperturbative intrinsic parton transverse momentum. The
scale dependent uncertainties of our calculations are about 20 \%
and they become larger at $q_T<2$ GeV, up to 70 \% at the $q_T\simeq
0$. However, the average value is still in agreement with
experimental data.

The UA1 Collaboration \cite{UA1} data for the $q_T-$spectrum of
Drell-Yan lepton pairs at $\sqrt{S}=630$ GeV are presented as a
differential invariant cross section, averaged over the virtual
photon mass in the range of $2m_\mu<Q<2.5$ GeV, and in the rapidity
range $|y|<1.7$. Because the cross section grows steeply when $Q\to
0$ and the lower boundary in $Q$ is fixed at the minimal kinematical
value $Q_{min}=2m_{\mu}$ , we need to take into account muon mass
$m_\mu$ in our calculations. Formally, it boils down to an
additional threshold factor $(1-4 m_\mu^2/Q^2)^{3/2}$ in a formula
for the $q_T-$spectrum.  Taking into account the large experimental
error bars, we can conclude that our calculations agree with the
data too.

\begin{figure}[ht]
\begin{center}
\includegraphics[width=.6\textwidth, clip=]{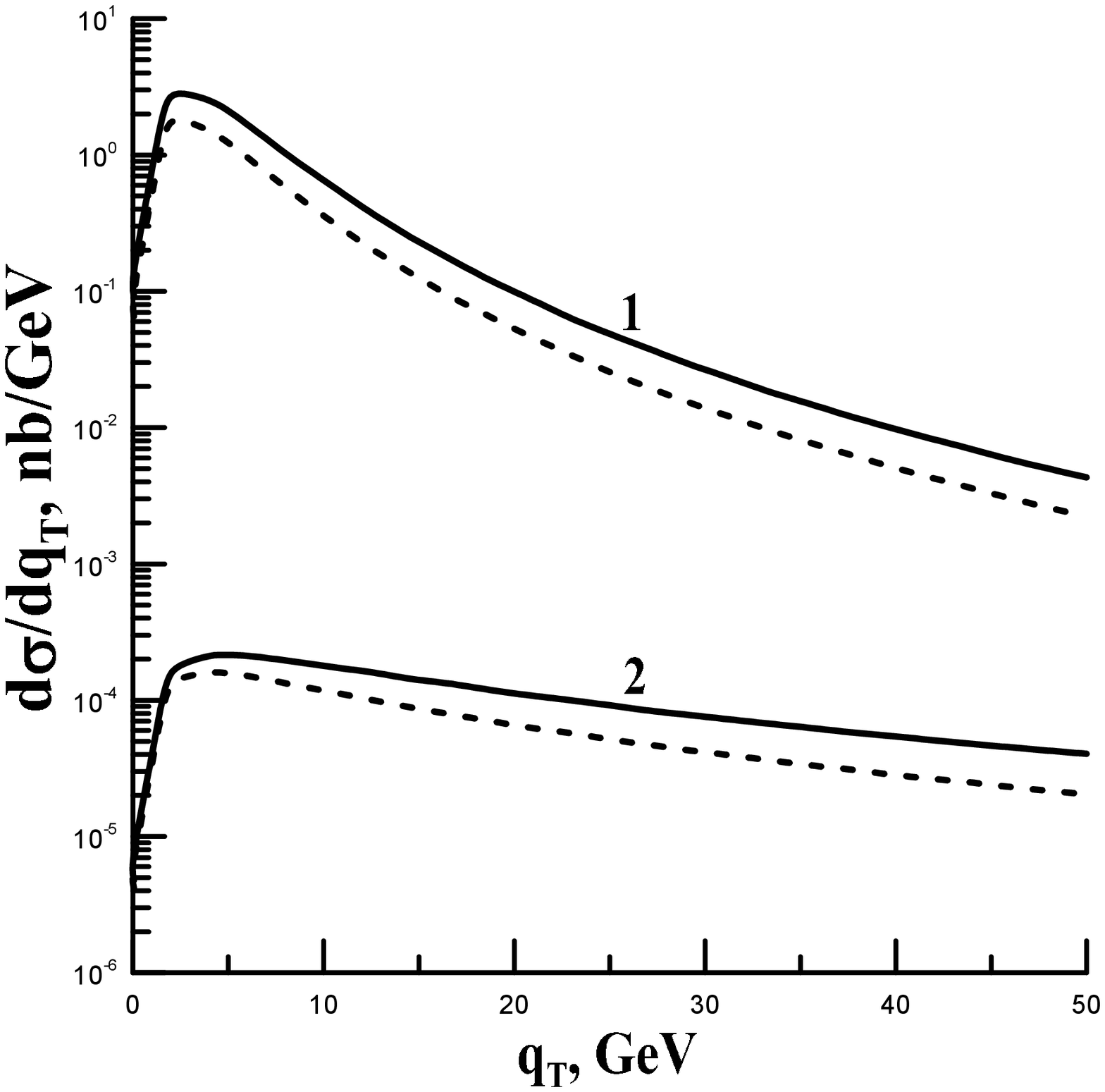}
\end{center}
\caption{ Differential cross section of Drell-Yan lepton pair
production as function of $q_T$ at the CERN LHC energy $\sqrt{S}=7$
TeV, and $|y|<3$. Curve 1 - $5<Q<50$ GeV, curve 2 - $120<Q<200$
GeV.}\label{LHCqt}
\end{figure}

In  Fig.~\ref{LHCqt}, we plot our predictions for the
transverse-momentum spectra of Drell-Yan lepton pairs at the CERN
LHC for energies $\sqrt{S}=7$ and $\sqrt{S}=14$ NeV in two regions
of virtual photon masses: $5<Q<50$ GeV, and $120<Q<200$ GeV. The
invariant-mass spectra have been obtained after integration over
rapidity in the range of $|y|<3$.

In the next section we compare our theoretical results  obtained in
the LO PRA with the experimental data for the angular coefficients
in Drell-Yan pair production. We consider data from the NuSea
Collaboration \cite{NuSea} at the Tevatron Collider, which
correspond to $\sqrt{S}=39$ GeV . We also make prediction for
angular coefficients $A_0,A_2$ for CERN LHC at energies $\sqrt{S}=7$
and $\sqrt{S}=14$ TeV.

\begin{figure}[p]
\begin{center}
\includegraphics[width=.6\textwidth, clip=]{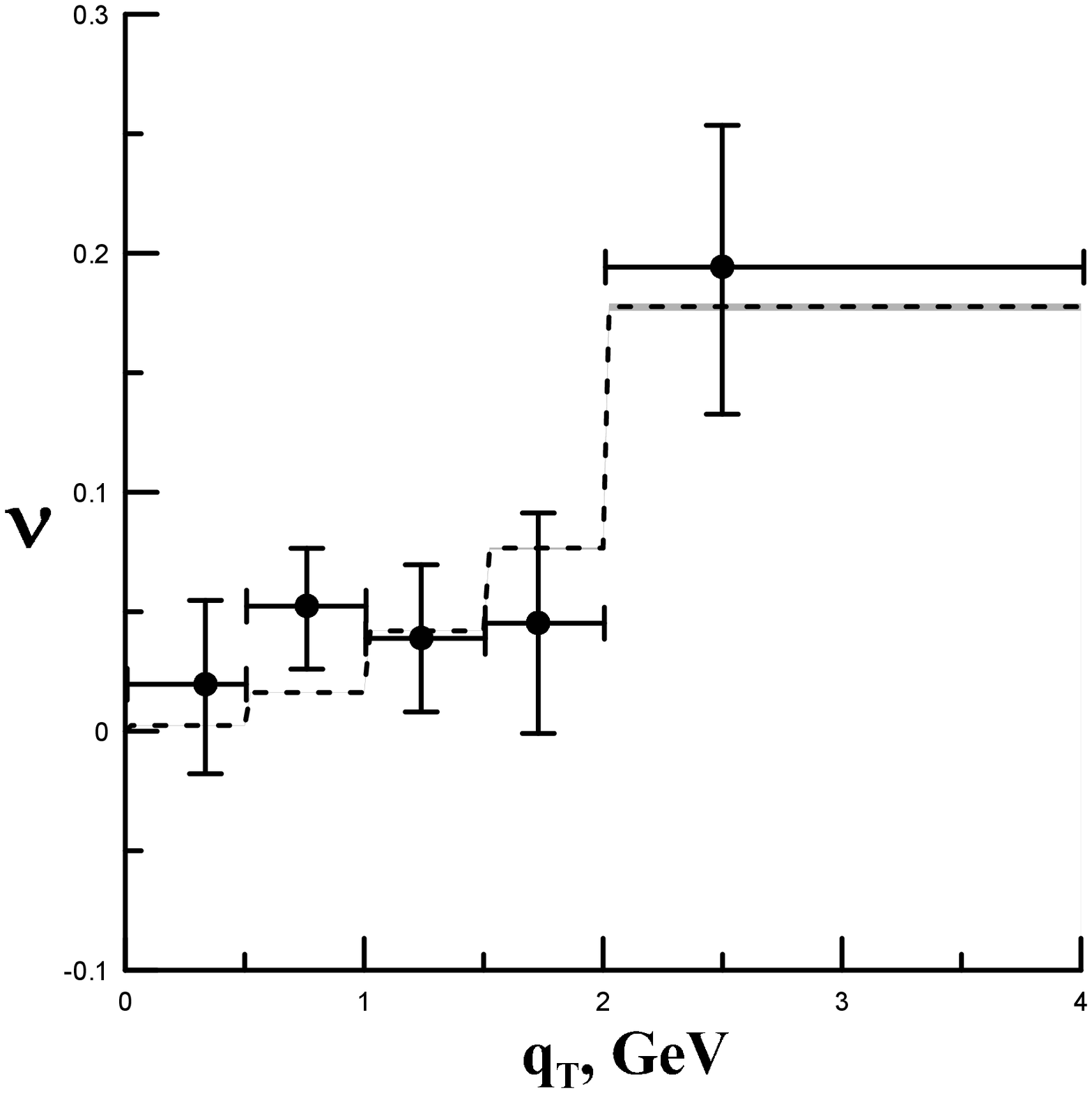}
\end{center}
\caption{Angular coefficient $\nu$ as function of $q_T$. The
histogram corresponds to LO calculation in  PRA with KMR \cite{KMR}
unintegrated PDFs. The data are from NuSea Collaboration
\cite{NuSea}.}\label{fig:nu}
\end{figure}

\begin{figure}[p]
\begin{center}
\includegraphics[width=.6\textwidth, clip=]{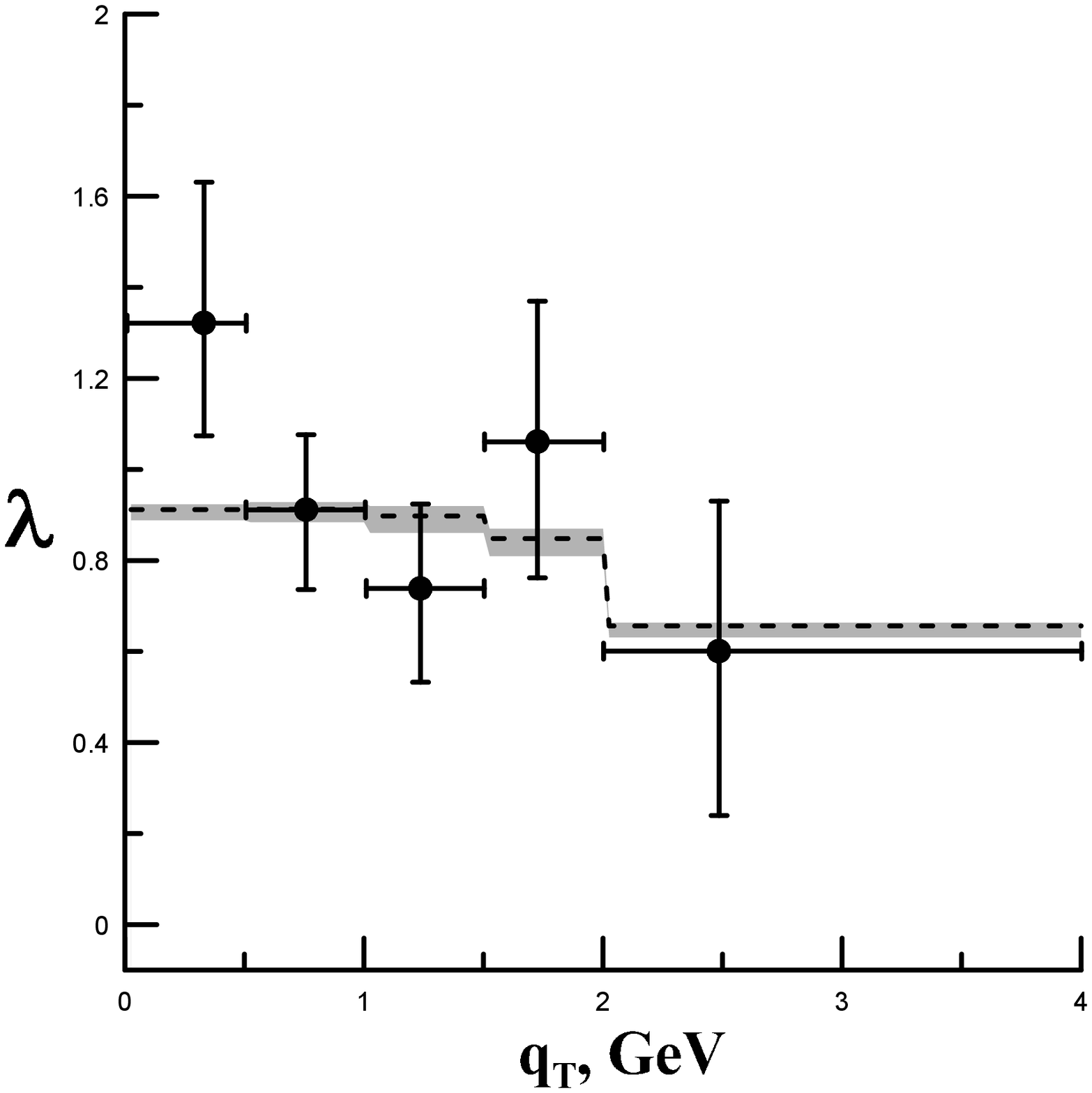}
\end{center}
\caption{ Angular coefficient $\lambda$ as function of $q_T$. The
histogram corresponds to LO calculation in  PRA with KMR \cite{KMR}
unintegrated PDFs. The data are from NuSea Collaboration
\cite{NuSea}.}\label{fig:lambda}
\end{figure}

NuSea Collaboration from Fermilab Tevatron recently has published
data \cite{NuSea} for Drell-Yan lepton pair production in
fixed-target experiment with hydrogen and deuterium targets and
$E_p=800$ GeV proton beam ($\sqrt{S}=39$ GeV). The measurements have
been done in the following kinematic domain: $4.5<Q<15$ GeV,
$0<q_T<4$ GeV, $0<x_F<0.8$. The results of measurements of angular
distributions are presented in terms of angular coefficients
$\lambda, \nu, \mu$ as functions of virtual photon transverse
momentum. We find good agreement of our LO PRA calculations with
data for $\nu$ and $\lambda$ at all values of $q_T$, as it is shown
in Figs.~\ref{fig:nu} and \ref{fig:lambda}. Additionally, we predict
$\mu=0$ which is also in agreement with the data within the
experimental error bars.

\begin{figure}[p]
\begin{center}
\includegraphics[width=.6\textwidth, clip=]{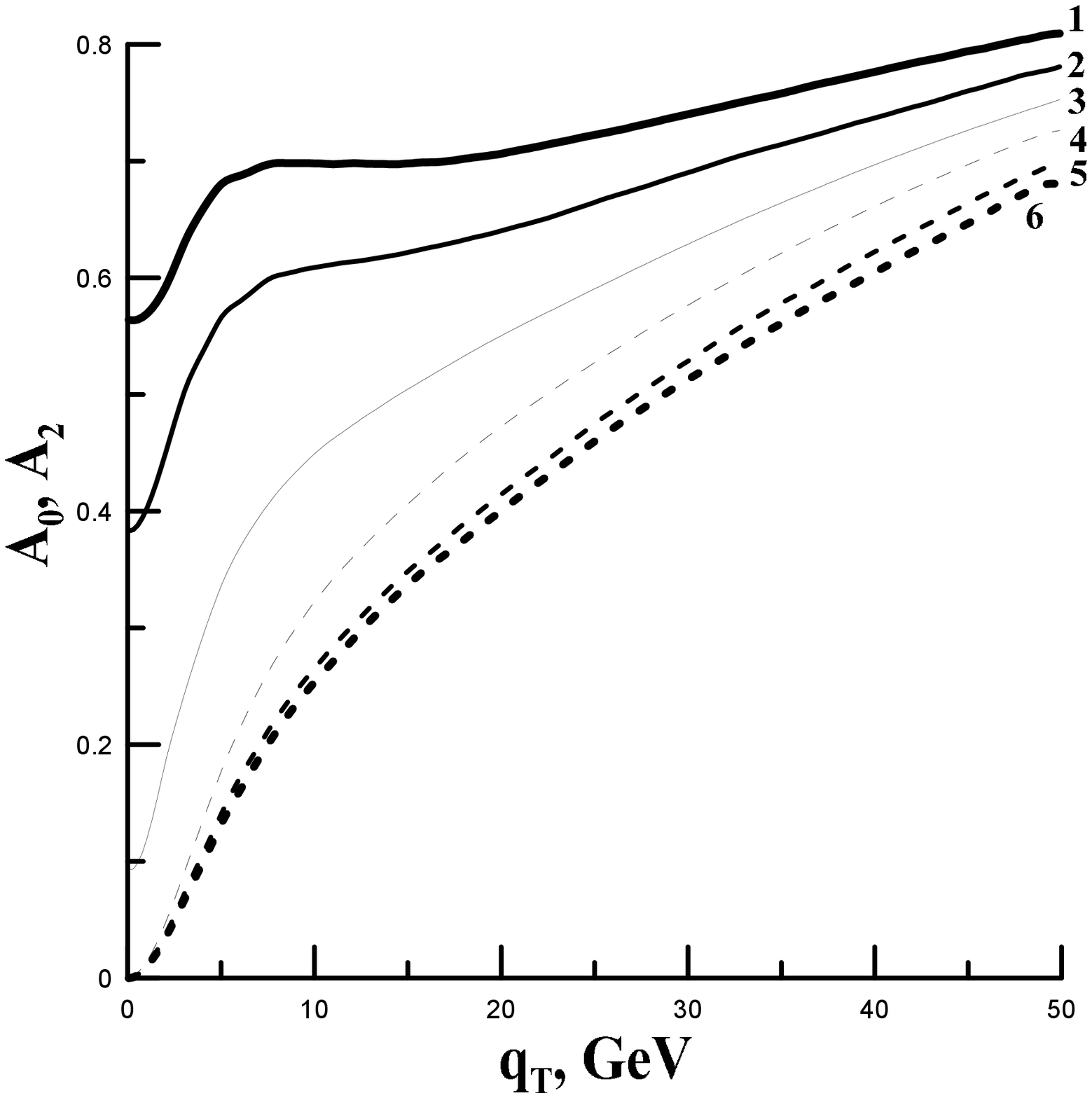}
\end{center}
\caption{Angular coefficients $A_0$ (solid curves) and $A_2$ (dashed
curves) in the proton-proton collisions
 as functions of $q_T$ for the mass region $5<Q<50$ GeV and different center of mass energies. Curves 1,6 -- 14 TeV,
  2,5 -- 7 TeV, 3,4 -- 2 TeV.}\label{fig:A0A2S1}
\end{figure}

\begin{figure}[p]
\begin{center}
\includegraphics[width=.6\textwidth, clip=]{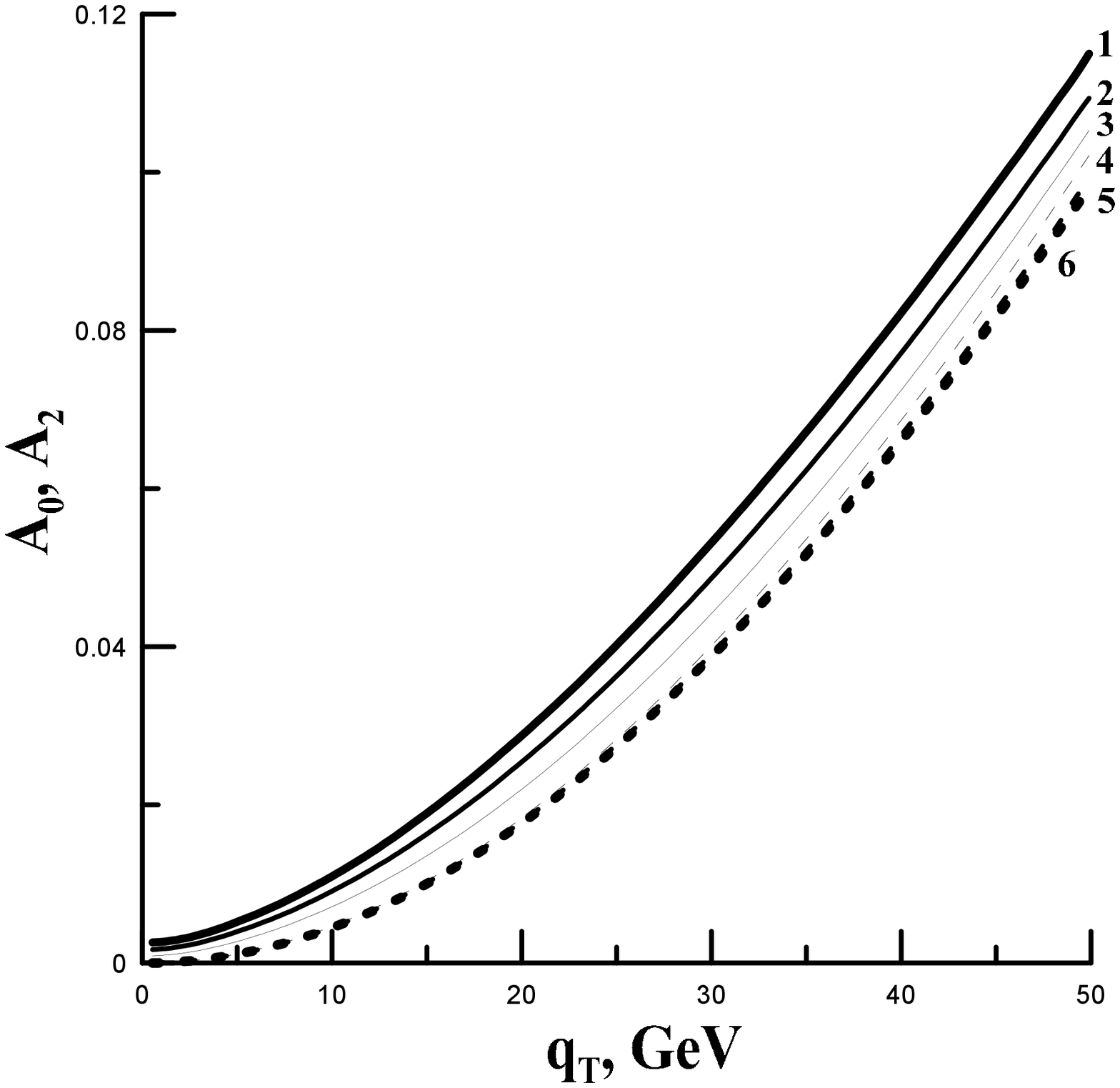}
\end{center}
\caption{ Angular coefficients $A_0$(solid curves) and $A_2$(dashed curves) in the proton-proton collisions
 as functions of $q_T$ for the mass region $120<Q<200$ GeV and different center of mass
 energies.  Curves 1,6 -- 14 TeV,
  2,5 -- 7 TeV, 3,4 -- 2 TeV.}\label{fig:A0A2S2}
\end{figure}

{
  It is known since the works of Lam and Tung \cite{LamTungRel}, that
  with allowance for  parton subprocesses (\ref{ann}) and (\ref{comp}) in the NLO collinear parton model,
   one can obtain the relation for the angular coefficients,
  $A_0 \simeq A_2$.  This relation is known to be valid at large $q_T$, and it
  has been
  verified experimentally in the Z-boson resonance mass region in \cite{CDF_A0_A2}.
  For the region of small dilepton masses it has not been verified yet.
  As it can be seen from the Figures \ref{fig:A0A2S1} and \ref{fig:A0A2S2},
  in our approach, this relation is approximately valid for energies below 1 TeV and for the large-mass region.
   For  large energies, and especially for the small-mass region (see Fig. \ref{fig:A0A2S1}),
   this relation is broken at the small values of $q_T$. Using the formulas (\ref{wTLPRA}),
   (\ref{wDPRA}),(\ref{WPRA})
   and definitions of angular coefficients (\ref{AAA}), one can show that:
  \begin{eqnarray}
  A_0(S,Q^2,y,q_T=0)=\frac{\sum\limits_q\int dt \Phi_q^p(x_1,t)
  \Phi_{\bar{q}}^p(x_2,t)\times 4t}{\sum\limits_q\int dt \Phi_q^p(x_1,t)\Phi_{\bar{q}}^p(x_2,t)\times (Q^2+2t)}, \\
  A_2(S,Q^2,y,q_T=0)=0,
  \end{eqnarray}
  where $t=t_1=t_2$, because of  $\vec {q}_{1T}=-\vec
{q}_{2T}$ if $\vec {q}_T=0$.
  So, the value of the coefficient $A_0$ at the $q_T=0$ characterizes the smearing of unintegrated PDF in transverse momentum
   and it is increasing
in the small $x\sim Q_T/\sqrt{S}$ region as it shown at the Fig.
\ref{fig:A0A2S1}.
  }

\section{Conclusions}
We reported a study of the Drell-Yan lepton pair production at LO in
the Parton Reggeization Approach, including subprocess~(\ref{QQDY})
with Reggeized quarks in the initial state. The Reggeization allows
to account in a simple and compact form the initial and final state
radiation effects with full allowance for finite transverse momenta
of partons. Our theoretical predictions provide an adequate
numerical description of a multitude of the experimental
measurements of lepton pair distributions on the invariant mass
($Q$), lepton pair transverse momentum ($q_T$) and longitudinal
scaling variable ($x_F$) as well as lepton pair angular
distributions at the SPS, Tevatron and LHC Colliders. This good
description is achieved, without any ad-hoc adjustments of input
parameters. By contrast, in the collinear Parton Model, such a
degree of agreement calls for NLO and NNLO corrections  and
complementary soft-gluon resummations and ad-hoc nonperturbative
transverse momenta of partons. In conclusion, the Parton
Raggeization approach has once again proven to be a powerful tool
for the theoretical description of QCD processes induced by
Reggeized gluons fusion as well as Reggeized quarks annihilation in
the high-energy limit.

\section*{Acknowledgements}

We are grateful to B.~A.~Kniehl and A.~V.~Shipilova for useful
discussions. The work was supported in part by the Ministry for
Science and Education of the Russian Federation under Contract
No.~14.B37.21.1182. The work of M.~N. is supported also by the Grant
of the Student's Stipend Program of the Dynasty Foundation. The work
of V.~S. was supported in part by the Russian Foundation for Basic
Research under Grant 11-02-00769-a and by SFB Fellowship of Hamburg
University (SFB-676). N.~N. acknowledges a support by Institut f.
Kernphysik, Forschungszentrum Juelich at the early stages of this
study.

\end{document}